\documentclass[aps,prl,preprint,preprintnumbers,amsmath,amssymb]{revtex4}

\usepackage{graphicx}
\usepackage[colorlinks=true,citecolor=blue]{hyperref}

\begin{document}
    %\linenumbers
    \title{Loss-enabled sub-Poissonian light generation in a bimodal nanocavity}
    \author{Arka Majumdar}
    \email{arkam@stanford.edu}
    \author{Michal Bajcsy}
    \author{Armand Rundquist}
    \author{Jelena Vu\v{c}kovi\'{c}}
    \affiliation{E.L.Ginzton Laboratory, Stanford University, Stanford, CA, $94305$\\}
    %\date{\today}
\begin{abstract}
We propose an implementation of a source of strongly sub-Poissonian light in a system consisting of a quantum dot coupled to both modes of a lossy bimodal optical cavity.
When one mode of the cavity is resonantly driven with coherent light, the system will act as an efficient photon number filter, and the transmitted light will have a strongly sub-Poissonian
character. In addition to numerical simulations demonstrating this effect, we present a physical explanation of the underlying mechanism. In particular, we show that the effect results from an interference between the coherent light transmitted through the resonant cavity and the super-Poissonian light generated by photon-induced tunneling.
Peculiarly, this effect vanishes in the absence of the cavity loss.
\end{abstract}
\maketitle
%\section{Introduction}
An optical cavity containing a strongly coupled quantum emitter, such as an atom or a quantum dot (QD), constitutes a system in which an optical nonlinearity is present even at a single photon level \cite{majumdar_pulsed, arka_switching, ScienceControlPhase}. The eigen-energies of this coupled system form an anharmonic ladder, which gives rise to phenomena like photon blockade and photon-induced tunneling \cite{birnbaum_nature, AF_natphys, AM_tunneling,blockade_imamog}. In photon blockade, coupling of a single photon to the system hinders the coupling of the subsequent photons, whereas in
photon-induced tunneling, coupling of initial photons favors the coupling of the subsequent photons. In an experiment, the signature of blockade or tunneling is observed by measuring the second order autocorrelation function $g^{(2)}(0)$; $g^{(2)}(0)<1$ $(>1)$ demonstrate the sub-Poissonian (super-Poissonian) photon statistics of the transmitted light and indicate that the system is in a photon blockade (tunneling) regime.

Photon blockade can be used to route photons in a quantum photonic circuit \cite{dayan_PRA}, or to mimic interacting bosons for efficient simulation of complex quantum phase transitions \cite{cca_plenio,cca_andrew, carusotto_imamoglu}. While most of the recent experiments focus on photon blockade with a single two level system and a single cavity \cite{birnbaum_nature, AF_natphys, AM_tunneling,blockade_imamog}, there have been several theoretical
proposals predicting photon blockade in systems based on multi-level atoms in a cavity \cite{97ImamogluBlockade}
and on a quantum dot interacting with a pair of proximity-coupled nanocavities  \cite{savona_ph_molecule, imma_ph_molecule} or wave-guides \cite{gerace_josephson_interfero}.

The cavity quantum electrodynamic (cQED) systems where photon blockade could be studied depend on three important rate
quantities: the coherent coupling strength between the atomic system and the cavity $g$, cavity field decay rate $\kappa$ and
the dipole decay rate $\gamma$. In all aforementioned proposals, the photon blockade occurs when the coherent interaction strength is
larger than the loss rates in the system. In fact, the limit of $g/\kappa,g/\gamma \rightarrow \infty$ results in vanishing overlap between the energy eigen-states of
the anharmonic ladder, which in turn leads to a perfect photon blockade ($g^{(2)}(0)=0$). In a solid
state optical system based on a photonic-crystal cavity with an embedded single QD as the two-level system, the condition $g\gg\gamma$ is generally easy to satisfy.
However, achieving the condition of $g\gg\kappa$, which requires a high quality (Q) factor of the cavity, is generally difficult due to fabrication challenges.
As a result, even the best photon blockade with a QD embedded in a solid-state nanocavity reported so far in the
literature gives a second order correlation $g^{(2)}(0)\sim0.75$ \cite{blockade_imamog}. Though a proposal based on a QD interacting with a photonic molecule
(a pair of coupled cavities) predicts efficient blockade even for cavities with easily achievable Q factors
\cite{savona_ph_molecule, imma_ph_molecule}, the suggested scheme requires
both individual addressability of each cavity and a large coupling strength between the two cavities. Since nanophotonic cavities are generally coupled via spatial
proximity, large coupling poses a major challenge for achieving individual addressability (see Appendix).

In this paper, we propose a different approach for generation of strongly anti-bunched light which
employs a bimodal cavity with both of its modes coupled to a QD. We will show that in this approach the cavity loss is actually crucial
for achieving the effect, as opposed to photon blockade systems introduced so far in which the cavity loss plays a negative role.
Specifically, the effect does not occur in our system in the limit of $g/\kappa
\rightarrow 0$, which is intuitively expected, as this is what also happens for the cases of blockade in a single cavity with strongly coupled QD and in a photonic molecule.  However, for $g/\kappa \rightarrow \infty$, the proposed system fails to generate sub-Poissonian light, in contrast with the single cavity with a strongly coupled QD, where a perfect photon blockade occurs in such limit. Here, we provide an intuitive explanation of how a balance between the coherent QD-cavity interaction and the decay of the cavity field is required to achieve strong sub-Poissonian output photon stream. Additionally, we analyze the nanophotonic platform for possible experimental
realization of this effect.

 %\section{Photon Blockade}

\begin{figure}
\centering
\includegraphics[width=3.5in]{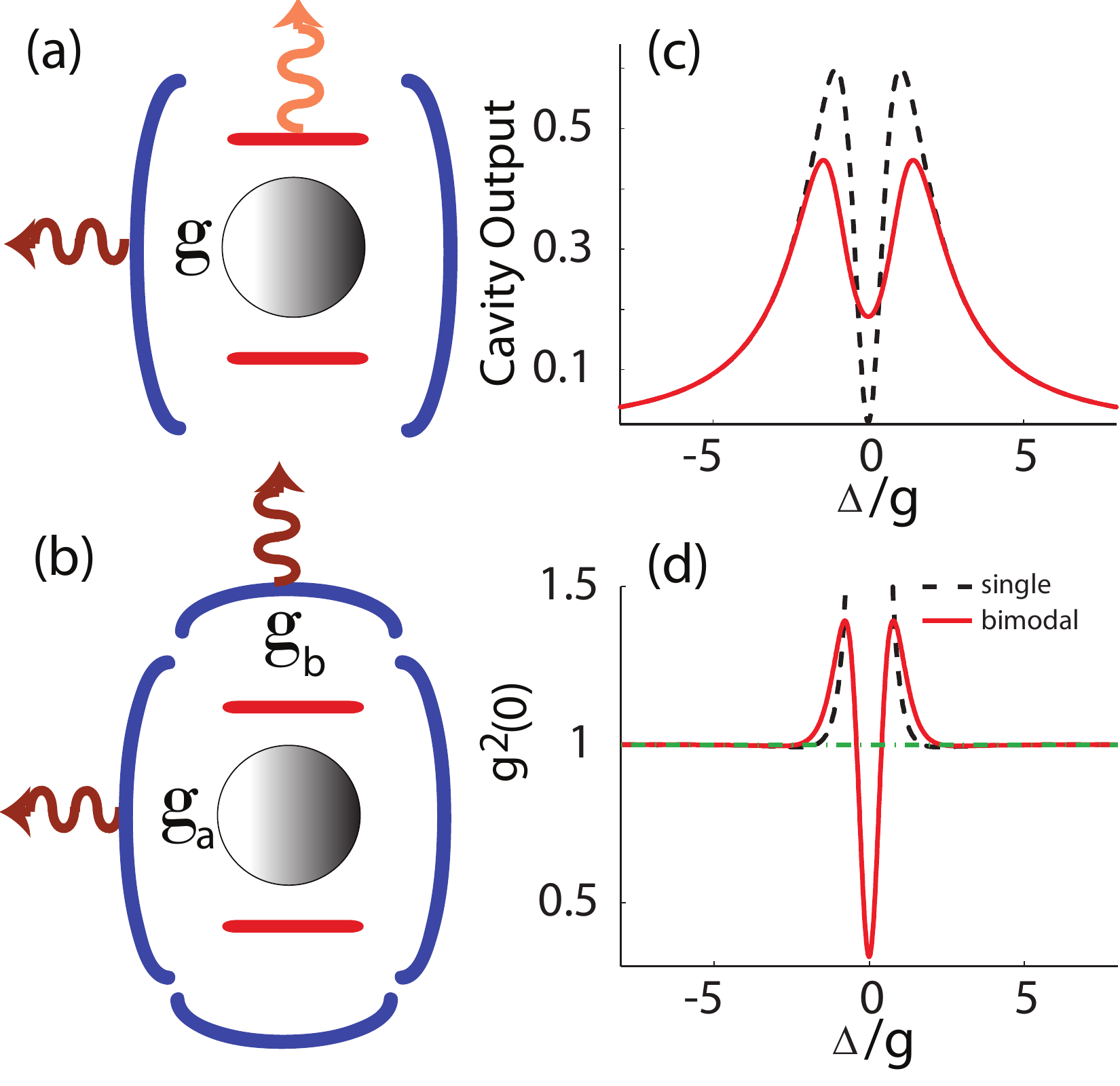}
\caption{(color online) (a) Schematic of a QD coupled to a single
mode cavity, with a coupling strength of $g$. (b) Schematic of a
bimodal cavity with a coupled QD. The two cavity modes are not
directly coupled to each other. However, both of them are coupled
to the QD with interaction strength $g_a$ and $g_b$. (c) The
cavity output $\kappa\langle a^\dag a\rangle$ as a function of the
driving laser detuning $\Delta$ from the empty cavity resonance
both for a single mode cavity (dashed line) and the bimodal cavity
(solid line). The split resonance observed is due to the coupled
QD. (d) Second order autocorrelation $g^{(2)}(0)$ function calculated for the collected
output of the driven mode for a single mode (dashed line) and bimodal cavity (solid line). The green dashed
line marks the Poissonian statistics of a coherent state. For
bimodal cavities we assumed identical interaction strength and
cavity decay rates for two modes. Parameters used for the
simulations: QD-cavity interaction strength $g/2\pi=g_a/2\pi=g_b/2\pi=10$ GHz,
cavity field decay rate $\kappa/2\pi=20$ GHz, dipole decay rate
$\gamma/2\pi=1$ GHz, and driving laser strength $\mathcal{E}/2\pi=1$
GHz.} \label{schematic_theory}
\end{figure}

In a conventional strongly coupled QD-cavity system, a QD interacts with a single cavity mode (Fig. \ref{schematic_theory}a). In a
bimodal cavity, the QD is coupled to both cavity modes (with photon annihilation operators $a$ and $b$) although there is no
direct coupling between the two modes (Fig. \ref{schematic_theory}b). Assuming the cavity modes are degenerate and the QD is resonant with both of them, the Hamiltonian $\mathcal{H}$ describing such a
system (in a frame rotating at the frequency of the laser driving the cavity mode $a$) is:
\begin{eqnarray}
\mathcal{H}=\Delta(a^\dag a&+&\sigma^\dag\sigma+b^\dag b)+g_a(a^\dag \sigma+a \sigma^\dag) \\ \nonumber
&+&g_b(b^\dag \sigma +b\sigma^\dag)+\mathcal{E}(a+a^\dag)
\end{eqnarray}
Here, $\sigma$ is the QD lowering operator, $g_a$ and $g_b$ are the coupling strengths between the QD and the two cavity modes,
$\mathcal{E}$ denotes strength of the driving laser and $\Delta$ is the detuning between the driving laser and the cavity modes. The loss in the
system is incorporated in the usual way by using the Master equation (see Appendix). The numerical calculations are
performed using the integration routines provided in the quantum optics toolbox \cite{qotoolbox}. Fig. \ref{schematic_theory}c
shows the transmitted light collected from the driven cavity ($\kappa\langle a^\dag a\rangle$) for both single (dashed line) and double mode
cavities (solid line). The cavity output is qualitatively similar for both cases, and the split resonance is caused by coupling of the QD to
the cavity and creation of polaritons. For the single mode cavity, the two polaritons are separated by $2g$, while for the bimodal cavity, the
separation is $2\sqrt{2}g$ due to the presence of two modes, as will be explained later. Increased cavity transmission at
$\Delta=0$ for the bimodal case is also due to the presence of two modes. However, the second-order autocorrelation function of the cavity
transmission $g^{(2)}(0)=\frac{\langle a^\dag a^\dag a a\rangle}{\langle a^\dag a\rangle^2}$ are strikingly different for two cases
(Fig. \ref{schematic_theory}d). For the single mode cavity, one observes photon blockade ($g^{(2)}(0)<1$), when the driving laser
is tuned to the frequency of the polariton, $\Delta\approx \pm g$. For the bimodal cavity, sub-Poissonian statistics is observed at three different detunings:
$\Delta\approx \pm \sqrt{2}g$ and $\Delta=0$. The weak sub-Poissonian light $(g^2(0)\sim 0.95)$ at $\Delta\approx \pm \sqrt{2}g$ is comparable to that observed in the single mode cavity,
and it arises from the same mechanism. At $\Delta=0$, the sub-Poissonian character is much stronger $(g^2(0)\sim 0.4)$, and it is this regime in the bimodal cavity that we will focus on.
Note that the sub-Poissonian character observed at this frequency of the driving laser cannot be explained by the anharmonic nature of the ladder alone. In fact, in the energy structure of the coupled QD and the bimodal cavity, we find an available state always at this empty cavity frequency (see Appendix).

\begin{figure}
\centering
\includegraphics[width=3.5in]{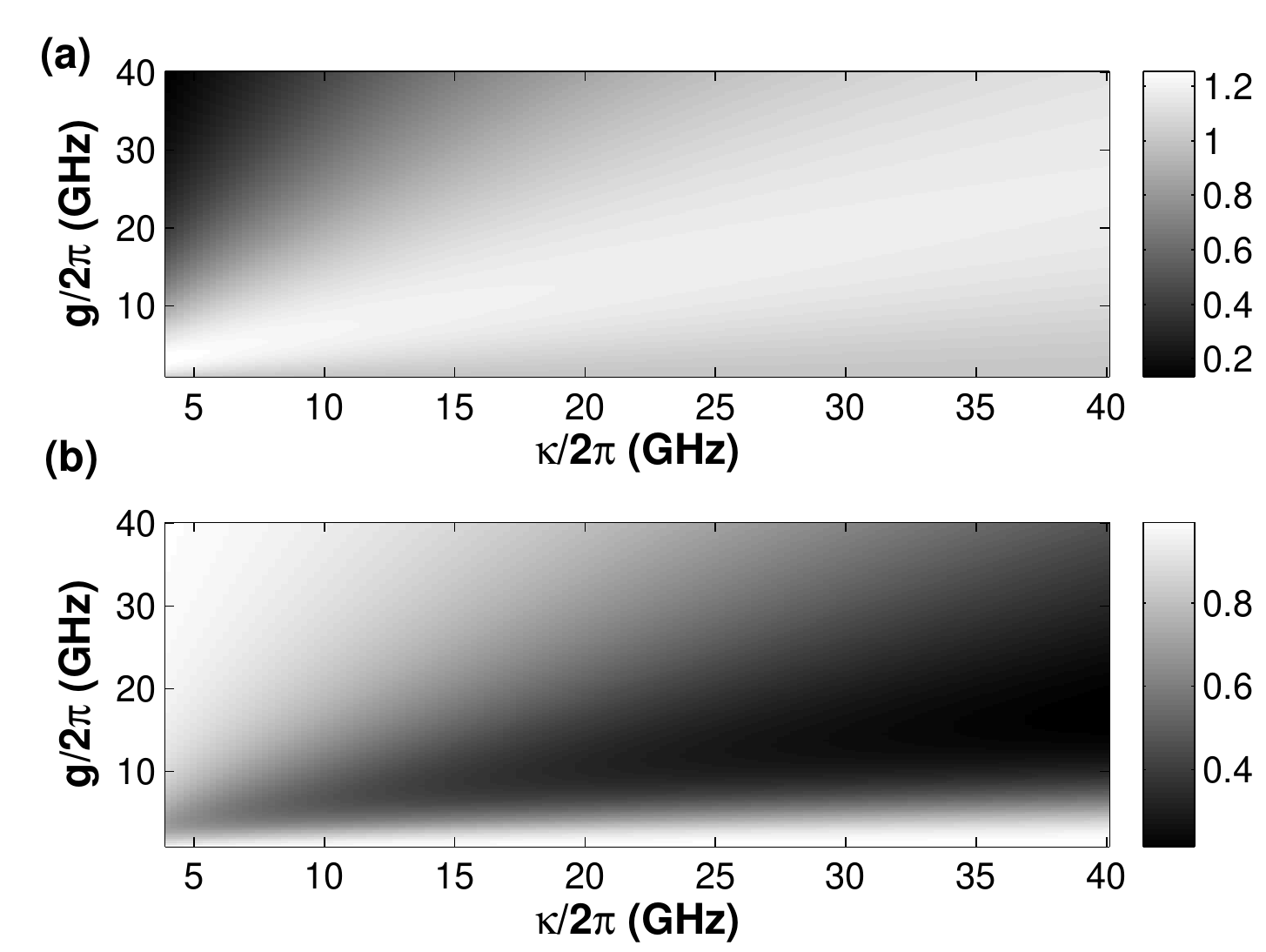}
\caption{(a) Second order autocorrelation $g^{(2)}(0)$ for the conventional photon blockade in a single
mode cavity as a function of the QD-cavity coupling strength $g$
and cavity field decay rate $\kappa$. $g^{(2)}(0)$
decreases with increasing value of  $g/\kappa$, as expected, as a result of reduced overlap of energy eigenstates in the anharmonic ladder. (b) $g^{(2)}(0)$ for the
bimodal cavity as a function of $g$ and $\kappa$. $g^2(0)$ is calculated for the output of the mode $a$, i.e., photons leaking from the mode $a$. We observe that
$g^{(2)}(0)$ tends to go to $1$ (Poissonian output) when $g/\kappa
\rightarrow 0$ or $\infty$. However, we can observe very low $g^{(2)}(0)$ even when
the QD is not strongly coupled to the two cavity modes ($g<\kappa/2$). }
\label{blockade_g_kappa}
\end{figure}

To further illustrate the difference between the photon blockade in a single mode cavity and the effect we observe in a bimodal cavity,
we perform numerical simulations for a range of coupling strengths $g$ and the cavity field decay rates $\kappa$ in both systems.
Using these simulations, we obtained the values of  $g^{(2)}(0)$ for the transmitted light
for a single mode cavity (laser tuned to one of the polaritons, i.e., $\Delta = g$) and for a double mode cavity (the laser tuned
to the bare cavity frequency, i.e., $\Delta=0$).
%For the simulations we assume $g_a/2\pi=g_b/2\pi=10$ GHz and the decay rate is $\kappa/2\pi=20$ GHz for both cavities.
Fig. \ref{blockade_g_kappa}a,b shows $g^{(2)}(0)$ as a function of $g$ and
$\kappa$. For a single mode cavity, blockade appears at high $g$
and low $\kappa$, as generally expected for any photon blockade
systems (Fig. \ref{blockade_g_kappa}a). However, for a bimodal
cavity, the effect disappears and the transmitted photon output returns to Poissonian whenever $g$ and $\kappa$ are
disproportionate (i.e., $g/\kappa\rightarrow 0$ or
$g/\kappa\rightarrow \infty$). A strongly sub-Poissonian output can be observed when $g$ and
$\kappa$ are comparable. Again, this result cannot be explained just
by the anharmonicity of the ladder of energy eigenstates.

\begin{figure}
\centering
\includegraphics[width=3.5in]{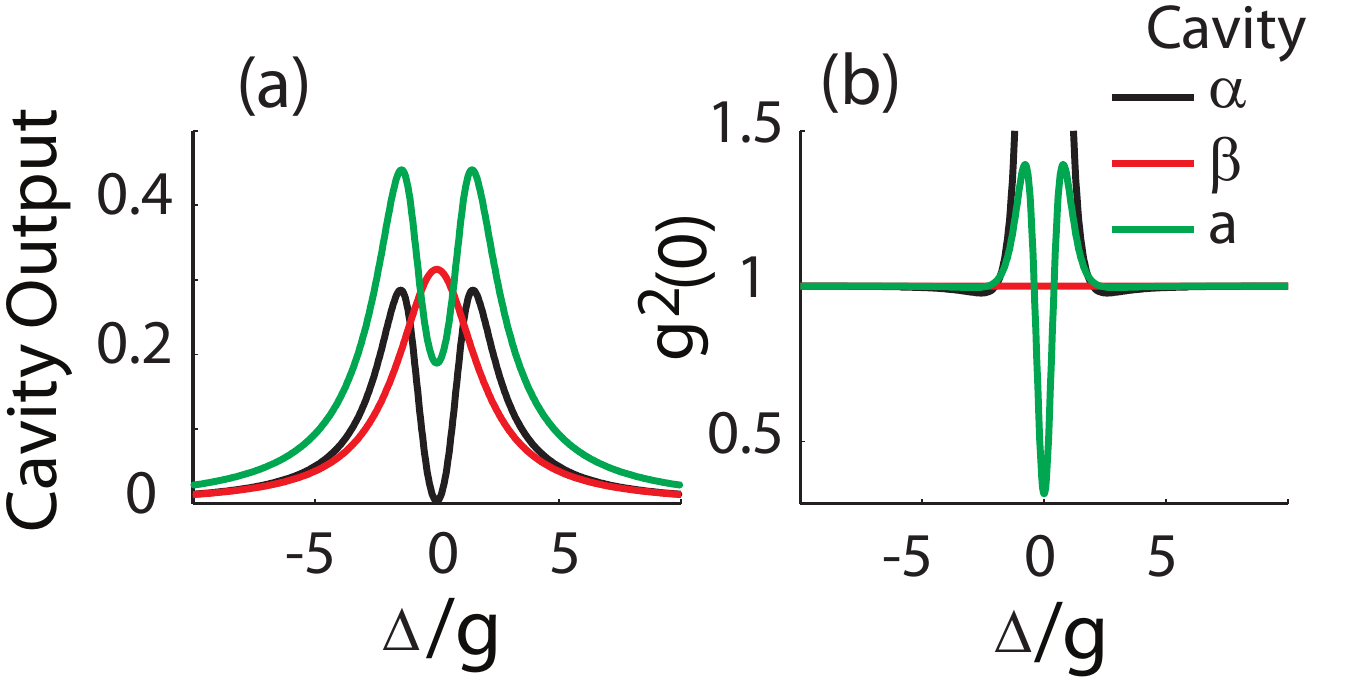}
\caption{(color online) (a) Cavity output for an empty cavity
$\beta$ and another cavity $\alpha$ coupled to a QD with a
coupling strength of $\sqrt{2}g$. The combined output of these two
produces the output from the bimodal cavity $a$. (b) $g^{(2)}(0)$ for these
three cases: the empty cavity $\beta$ gives Poissonian
light; the cavity $\alpha$ with coupled QD gives super-Poissonian
light due to photo-induced tunneling \cite{AM_tunneling} (the black curve goes to infinity at $\Delta=0$); the
combined output $a$ provides sub-Poissonian light. Parameters for the
simulation: $g/2\pi=10$ GHz, $\kappa/2\pi=20$ GHz, $\gamma/2\pi=1$
GHz and $\mathcal{E}/2\pi=1$ GHz.} \label{effective_model}
\end{figure}

%\section{Effective Model}
To understand the origin of the strongly sub-Poissonian light transmitted through a
bimodal cavity, we transform the system's Hamiltonian to a different cavity
mode basis: $\alpha=(a+b)/\sqrt{2}$ and $\beta=(a-b)/\sqrt{2}$. The
Hamiltonian $\mathcal{H}$ can be written (assuming $g_a=g_b=g$) as
$\mathcal{H}=\mathcal{H}_1+\mathcal{H}_2$ with
\begin{equation}
\mathcal{H}_1=\Delta (\alpha^\dag
\alpha+\sigma^\dag\sigma)+\sqrt{2}g(\alpha^\dag \sigma +\alpha
\sigma^\dag)+\frac{\mathcal{E}}{\sqrt{2}}(\alpha+\alpha^\dag)
\end{equation}
describing a driven single mode cavity coupled to a QD with a
strength of $\sqrt{2}g$ and
\begin{equation}
\mathcal{H}_2=\Delta \beta^\dag \beta+\frac{\mathcal{E}}{\sqrt{2}}(\beta+\beta^\dag)
\end{equation}
describing a driven empty cavity mode. Both cavities are
driven at the bare cavity resonances. We monitor
$a=(\alpha+\beta)/\sqrt{2}$ which, in the transformed basis, is equivalent to the output from two
cavities: one with a coupled QD ($\alpha$) and the other empty
($\beta$), combined in a beam-splitter. Fig.
\ref{effective_model}a shows the transmitted cavity output for
three different cases: cavity $\alpha$ alone, cavity $\beta$ alone
and the combined output. Note the polariton peaks in the combined
output at $\pm\sqrt{2}g$ and increased transmission of light at
zero detuning due to the empty cavity.

The cavity transmission with a strongly coupled QD driven at the cavity
resonance is super-Poissonian due to photon-induced tunneling
\cite{AM_tunneling} ($\alpha$ in Fig. \ref{effective_model}b). In this regime, the coupling of initial photons
into the system is inhibited by the absence of the dressed states at this frequency. However, once
the initial photon is coupled, the probability of coupling subsequent photons is increased as higher order manifolds in the ladder of dressed states are reached via multiphoton processes. In our system, as a result of broadening of the dressed states, at empty cavity resonance one can excite multiple higher order manifolds.
Hence, the light transmitted through a cavity in
the photon-induced tunneling regime is a superposition of Fock states with small photon numbers and a strong presence of the vacuum state. As a result,
the photon statistics of this light is super-Poissonian \cite{AM_tunneling}. On the other hand, the empty cavity
transmission ($\beta$ in Fig. \ref{effective_model} b) is a purely Poissonian coherent state. When the outputs of these two cavities are combined on a beam-splitter ($a=(\alpha+\beta)/\sqrt{2}$ in Fig. \ref{effective_model} b), the output shows sub-Poissonian character somewhat similar to the photon-added coherent states \cite{agarwal_photon_cohe,science_photon_cohe}.
However, for efficient generation of sub-Poissonian light, one needs comparable transmitted light intensity from both cavities, which calls for a balance between the cavity
loss $\kappa$ and QD-cavity nonlinear interaction strength $g$. Using this effective model, the somewhat unusual dependence of $g^{(2)}(0)$
on $g$ and $\kappa$ can now be explained. When $g/\kappa\rightarrow0$, the coupled system is linear and both of the equivalent cavities transmit just coherent light.
On the other hand, although photon-induced tunneling does happen in the limit $g/\kappa\rightarrow\infty$, the amount of
super-Poissonian light transmitted through the cavity $\alpha$ is so small (as the dressed states separation in the ladder is so large that it is impossible to couple photons at energies between them) that its interference with the coherent
light from the empty cavity $\beta$ will still result in light with Poissonian statistics.
%Thus to achieve a good blockade a balance between the nonlinearity and loss of the system is required.
To generate enough super-Poissonian light via photon-induced tunneling in cavity $\alpha$ to affect the coherent light from the empty cavity $\beta$, a moderate dot-cavity interaction strength $g$ is required.
%since the requirement on $g/\kappa$ for observing tunneling is less severe than the one needed for good blockade.
%Fig. \ref{effective_model}b then shows the $g^{(2)}(0)$ for the output from cavity $\alpha$, cavity $\beta$ and the combined
%output of those two.

%\section{Dependence on parameters}

Finally, we discuss the nanophotonic platform that can be
used to implement our proposal. A photonic-crystal cavity with $C_6$ symmetry
can support two degenerate cavity modes with orthogonal polarizations
\cite{calzone_immamoglu}. The two cavity modes are thus not coupled to each other (since their polarizations are orthogonal),
and can be easily addressed independently by a laser.
%On the other hand, a QD has orthogonal polarizations for its two exciton lines.
At the same time, a QD can be coupled to both cavity modes, if it is placed
spatially at the center of the cavity with its dipole moment aligned at $45^o$-angle to the polarizations of both modes.

\begin{figure}
\centering
\includegraphics[width=3.5in]{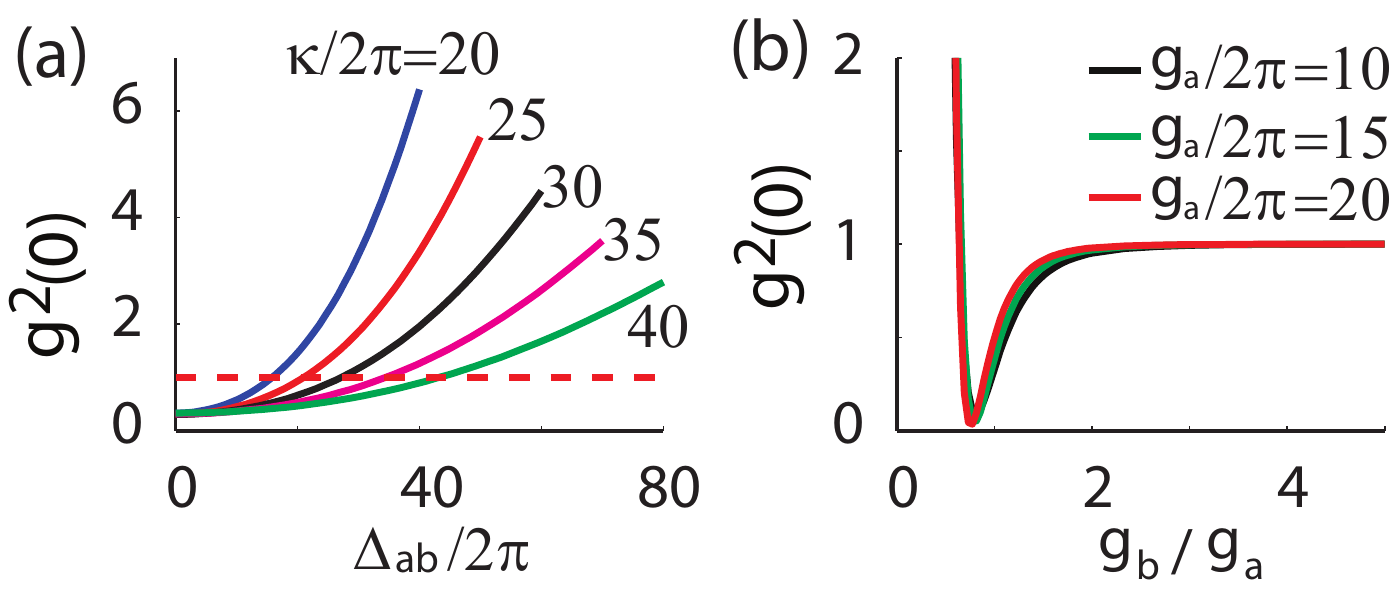}
\caption{(color online)(a) Second order autocorrelation $g^{(2)}(0)$
of the cavity transmission, as a function of the relative detuning
$\Delta_{ab}$ between two cavity modes for different cavity field
decay rate $\kappa=\kappa_a=\kappa_b$. The quality of the sub-Poissonian photon stream in the transmitted output degrades with
increasing detuning, which can be compensated by increasing
$\kappa$, while maintaining low $g^{(2)}(0)$. For these simulations we assume $g_a/2\pi=g_b/2\pi=10$ GHz.
(b) Second order
autocorrelation $g^{(2)}(0)$ of the cavity transmission as a function
of the ratio $g_b/g_a$ for different $g_a$. The transmitted light
behaves like a coherent state at high $g_b/g_a$ ratio and like a
super-Poissonian state generated via photo-induced tunneling at
low $g_b/g_a$ ratio. In between, when $g_b/g_a \sim 1$, we observe
strong sub-Poissonian output. Here, $\kappa/2\pi=20$ GHz for both cavity modes. } \label{Param_blockade}
\end{figure}

Two potential issues can arise from fabrication imperfections in a realistic system:
a frequency difference $\Delta_{ab}$ between the two cavity modes and a mismatch between the QD coupling strengths $g_a$ and
$g_b$ to each mode. These issues can be seen in the preliminary experimental results shown in the Appendix. To examine the robustness of the proposed scheme against these imperfections, we plot their effects on $g^{(2)}(0)$ in Fig. \ref{Param_blockade}.
Fig. \ref{Param_blockade}a shows the numerically calculated $g^{(2)}(0)$ as a function of the detuning $\Delta_{ab}$. We observe that the sub-Poissonian character of the transmitted light vanishes when
$\Delta_{ab}\geq \kappa$. This negative effect of frequency difference of the two modes can be balanced simply by increasing the cavity decay rate $\kappa$,
i.e., by lowering the cavity quality factor. This results in an increase of the frequency overlap between the two modes and makes the degeneracy of the two modes more robust.
The effects of this improvement outweigh the penalty incurred on the system's performance by reducing the $g \over \kappa$ ratio, and we can see in Fig. \ref{Param_blockade}a that  and a strongly anti-bunched output can still be produced.
%This is not completely surprising since the photon blockade in a bimodal cavity happens at a very different $g/\kappa$ regime (Fig. \ref{blockade_g_kappa}b).
Additionally, we analyze the performance of the system as a function of the ratio $g_b/g_a$, where $g_b$ and $g_a$ are
the QD coupling strengths with the cavity modes $a$ and $b$ assuming mode $a$ is coherently driven. It can be shown from the effective
model that at a large $g_b/g_a$ ratio, we essentially drive only the empty cavity $\beta$ and the photon statistics are Poissonian.
Similarly, at a small $g_b/g_a$ ratio, we drive only the cavity $\alpha$ with coupled QD and the photon statistics are
super-Poissonian due to photo-induced tunneling (see Appendix). When $g_b/g_a \sim 1$, we meet the optimal condition
of interference between the coherent state and super-Poissonian state to generate light with sub-Poissonian
photon statistics. This can be seen in the numerical simulations of $g^{(2)}(0)$ as a function of $g_b/g_a$ in Fig.
\ref{Param_blockade}b. The system performance is insensitive to the actual value of $g_a$ for a relatively large range, as long as
the ratio $g_b/g_a$ is maintained. At the same time, we can see that the lowest value of $g^2(0)$ is achieved for the ratio of coupling strengths of $g_b/g_a \sim 0.8$. We note that, this ratio depends on the driving strength of the laser, and can be related to the requirement of the similar cavity transmission from the cavities $\alpha$ and $\beta$.

%\section{Experimental Feasibility: compare photonic molecule with this (supplement)+mention many body+show SEM; phase %bistability}
In summary, we introduced a scheme for generation of sub-Poissonian light in a cQED system with a bimodal cavity and provided a theoretical and numerical analysis of its performance.
For similar system parameters, the bimodal cavity can provide much better sub-Poissonian character of the transmitted photon stream ($g^{(2)}(0)\to 0$)
compared to a single mode cavity ($g^{(2)}(0)\sim 0.9$). The proposed effect happens due to interference
between a coherent state and a super-Poissonian state generated by photon-induced tunneling, and a balance between the
nonlinearity and the loss of the system is required to observe it. Moreover, the effect disappears in the absence of the cavity loss $(g/\kappa\rightarrow\infty)$. This interplay between loss and nonlinearity has a great potential to be
exploited for design of realistic coupled cavity arrays for efficient quantum simulation.

The authors acknowledge financial support provided by DARPA, ONR,
NSF and the ARO. A.R. is also supported by a Stanford Graduate Fellowship.

\appendix
\section{Eigen-structure of the quantum dot coupled to a bimodal cavity with loss:}
The Hamiltonian $\mathcal{H}$ describing the coupled system is
given by
\begin{equation}
\mathcal{H}=\Delta (a^\dag a+\sigma^\dag\sigma+ b^\dag
b)+g_a(a^\dag \sigma +a \sigma^\dag)+g_b(b^\dag \sigma +b
\sigma^\dag)+\mathcal{E}(a+a^\dag)
\end{equation}
To incorporate the loss in the system, we use a Master equation
approach. The lossy dynamics of the quantum system are governed by
\begin{equation}
\label{Maseq} \frac{d\rho}{dt}=-i[H,\rho]+ 2\kappa_a
\mathcal{L}[a]+2\kappa_b \mathcal{L}[b]+2\gamma
\mathcal{L}[\sigma].
\end{equation}
Here, $\rho$ is the density matrix of the coupled QD-cavity
system, $2\gamma$, $2\kappa_a$ and $2\kappa_b$ are the QD
spontaneous emission rate and the cavity population decay rates,
and we neglect any non-radiative decay of the QD
exciton. Finally, $\mathcal{L}[D]= D\rho D^\dag-\frac{1}{2}D^\dag D
\rho-\frac{1}{2}\rho D^\dag D$ is the Lindblad operator
corresponding to a collapse operator $D$.

Although the lossy dynamics describes the system in a more realistic
way, a great deal of information can be obtained just from
the coherent dynamics of the system, especially from the
eigen-structure of the combined system. In fact, the explanation
of photon blockade in the usual case of a two level system coupled
to a single mode cavity relies mostly on the anharmonic ladder
structure. This system is well-described by the Jaynes Cummings
Hamiltonian, and the energy eigenstates are grouped in two-level
manifolds with eigen-energies given by $ \pm g\sqrt{n}$ (for a
resonant dot-cavity system), where $n$ is the number of energy
quanta in the coupled QD-cavity system. The eigenstates can be
written as:
\begin{equation}
|n,\pm\rangle=\frac{|g,n\rangle \pm |e,n-1\rangle}{\sqrt{2}}
\end{equation}
This anharmonicity gives rise to photon blockade when the driving
laser is tuned to the polariton frequencies. In a similar
manner, one can diagonalize the Hamiltonian for the bimodal cavity
QED system. However, in contrast to the Jaynes-Cummings
Hamiltonian, the matrix size for a specific number of quanta
present in the bimodal cavity system is not fixed. In fact, for
$n$ quanta, the total number of bare basis states is $2n+1$. In
the Jaynes-Cummings Hamiltonian, the number of basis states is
always $2$, namely $|e,n-1\rangle$ and $|g,n\rangle$ for $n$
quanta present in the system. From the symmetry of the coupled
system, we find that the summation of the eigen-energies of the
matrix (trace of the matrix) with $n$ quanta is zero. As there is
an odd number of eigen-values, by symmetry, we conclude that there
is always an eigen-energy of $0$. Hence there is always an
eigen-state at the empty cavity resonance. So if the system is
driven resonantly, we should not expect any photon blockade caused
only by the anharmonicity of the ladder.

We will now describe the effective model, which is described in the
main text only for $g_a=g_b$ case. To understand the nature of the
effect resulting in transmission of sub-Poissonian light, we make the following transformation:
\begin{eqnarray}
  \alpha &=& \frac{g_a a+g_b b}{\sqrt{g_a^2+g_b^2}} \\
  \beta &=& \frac{g_b a-g_a b}{\sqrt{g_a^2+g_b^2}}
\end{eqnarray}
Under this transformation, we can rewrite the total Hamiltonian as
\begin{equation}
\mathcal{H}=\mathcal{H}_1+\mathcal{H}_2
\end{equation}
with
\begin{eqnarray}
  \mathcal{H}_1 &=& \Delta(\alpha^\dag\alpha+\sigma^\dag\sigma)+\sqrt{g_a^2+g_b^2}(\alpha\sigma^\dag +\alpha^\dag \sigma)+\frac{\mathcal{E}}{\sqrt{1+r^2}}(\alpha+\alpha^\dag) \\
  \mathcal{H}_2 &=& \Delta\beta^\dag\beta+\frac{r\mathcal{E}}{\sqrt{1+r^2}}(\beta+\beta^\dag),
\end{eqnarray}
where $r=g_b/g_a$. For $g_a=g_b$, we recover the result reported
in the main text. When $r \rightarrow 0$,
there is no light in the uncoupled cavity, and all the light
output is from the coupled QD-cavity system. That light is
super-Poissonian due to photo-induced tunneling. On the other
hand, when $r \rightarrow \infty$, the coupled QD-cavity does not
get any light. Hence the output light is just in a coherent state,
which means its photon statistics will be Poissonian.

\section{Light from the undriven cavity mode:}
\begin{figure}
\centering
\includegraphics[width=5in]{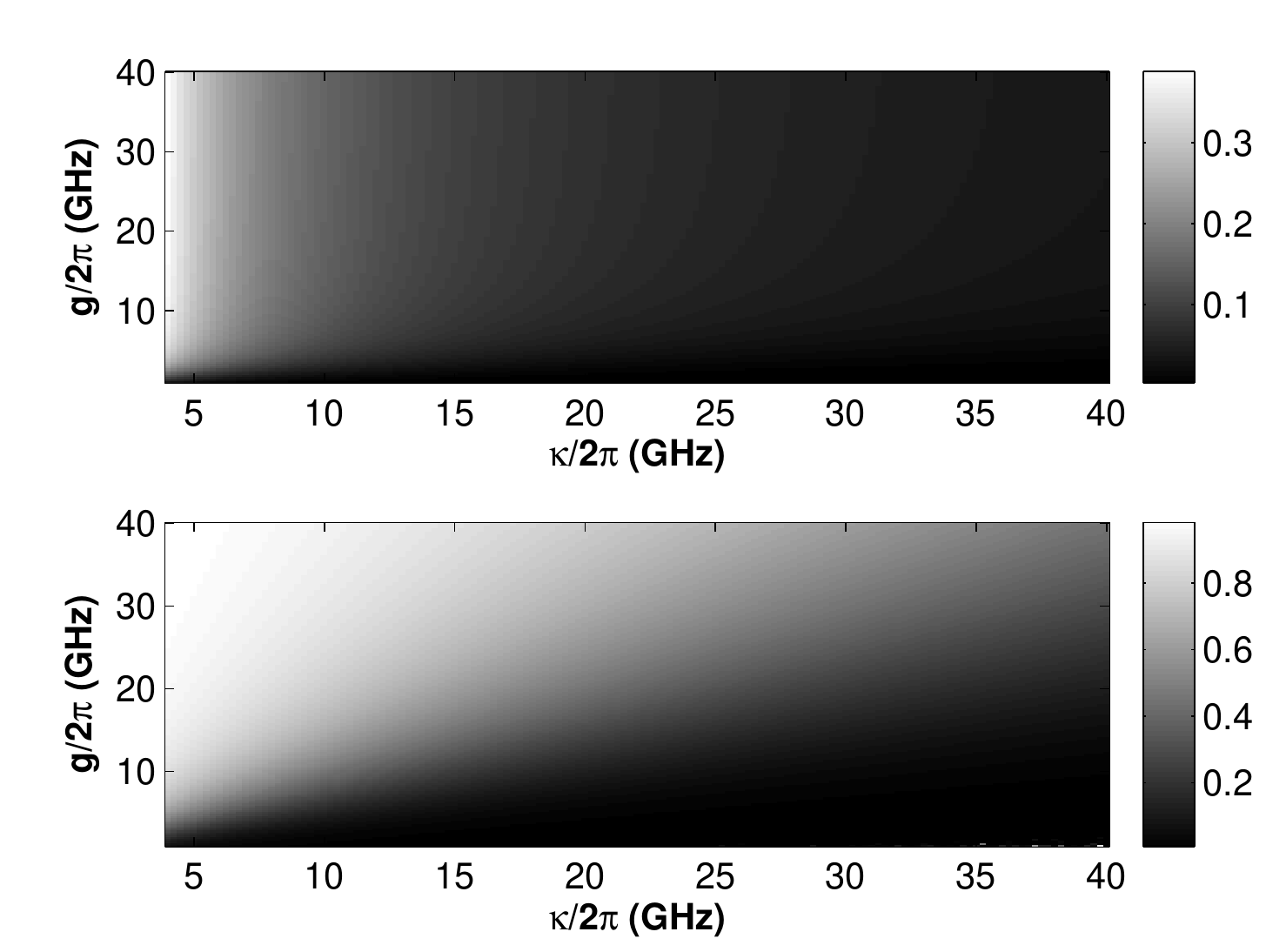}
\caption{(a) Transmitted output $\kappa \langle
b^\dag b\rangle$ from the undriven cavity mode $b$ as a function of $g$ and
$\kappa$. For both the modes same $g$ and $\kappa$ are assumed.(b)
Second order auto-correlation $g^2(0)$ for the bimodal cavity as
a function of $g$ and $\kappa$. The light is always
sub-Poissonian, and results form the resonant QD excitation via mode $a$, and collection of the output photons  via mode $b$. This is similar to a conventional single photon source under resonant excitation of the QD. } \label{modeb_light}
\end{figure}

Here, we briefly discuss the light state from the undriven
cavity mode $b$. Mode $b$ is not directly coupled to mode $a$, and
the only way it can get light is via the QD. Hence the amount of
light collected from mode $b$ is very small, but the light is
sub-Poissonian. However, there is no photon blockade present in
the mode $b$, and the sub-Poissonian character of light in mode $b$ is similar
to the single photons generated by resonant excitation of a QD.
Fig. \ref{modeb_light} a,b shows the transmitted cavity output and
second order auto-correlation $g^2(0)$ as a function of $g$ and
$\kappa$.

\section{Comparison with the photonic molecule coupled to a
quantum dot:} 
\begin{figure}
\centering
\includegraphics[width=5in]{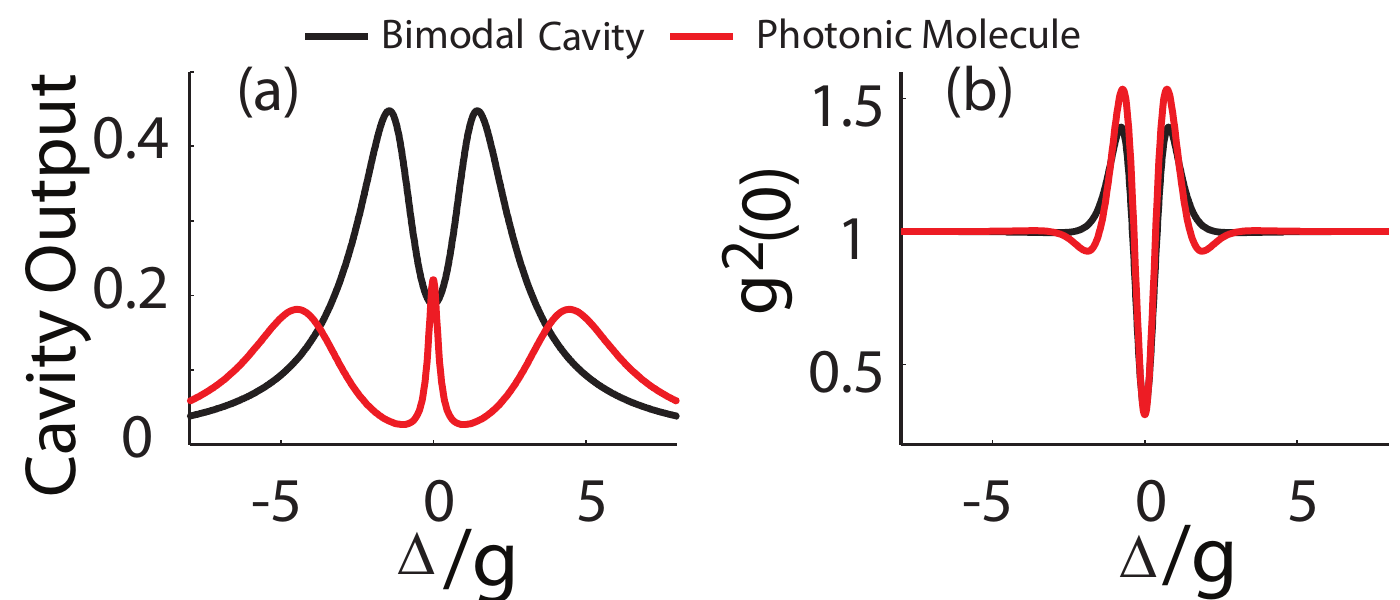}
\caption{(color online) (a) Cavity transmitted output for two
cases: bimodal cQED and photonic molecule. (b) Second order
autocorrelation $g^2(0)$ for those two cases. Parameters used for the simulations: QD-cavity interaction strengths $g/2\pi=g_a/2\pi=g_b/2\pi=10$ GHz;
cavity field decay rates $\kappa_a/2\pi=\kappa/2\pi=20$ GHz; coupling strength for the Photonic molecule $J/2\pi=40$ GHz; dipole decay rate $\gamma/2\pi=1$ GHz.}
\label{bimodal_phot_mol}
\end{figure}

Our method for is somewhat related
to the recent proposals of generation of sub-Poissonian light in a photonic molecule
\cite{savona_ph_molecule, imma_ph_molecule}. In that approach, one
needs two cavities coupled to each other, with one of the cavities
containing a weakly coupled QD. The empty cavity is driven, and
light is also collected from the empty cavity. Similarly to our bimodal
cavity proposal, sub-Poissonian light is obtained when the driving laser is
resonant with the system. The Hamiltonian $\mathcal{H}$ for such a photonic molecule is
\begin{equation}
\mathcal{H}=\Delta (a^\dag a+\sigma^\dag\sigma+ b^\dag
b)+g(b^\dag \sigma +b \sigma^\dag)+J(a^\dag b +ab^\dag)+\mathcal{E}(a+a^\dag)
\end{equation}
where $J$ is the cavity-cavity interaction strength. Figure
\ref{bimodal_phot_mol} a,b shows the transmitted cavity output and
$g^2(0)$ for two different cases: photonic molecule and the
bimodal cavity.
We observe that the performance for both the cases are comparable.
However, for photonic molecule, we need to maintain both
individual addressability and high coupling strength between the
cavities. In the nano-photonics platform this requirement is
challenging, as generally the inter-cavity coupling is achieved by
spatial proximity. Although, a wave-guide coupling can be
employed to mitigate the problem, that solution brings added complexity of
engineering the cavity-to-waveguide coupling. In addition, the cavity
output drops if the inter-cavity coupling strength is high, making
the photon statistics measurement challenging. On the other hand,
our proposal within a bimodal cavity does not require
any cavity-cavity coupling. For a degenerate photonic crystal
cavity, the two cavity modes are of opposite polarization, and
have no coupling. Furthermore, due to orthogonality of their polarizations, each cavity
mode can be easily addressed separately as required by the proposal.

\section{Preliminary experimental result on bimodal photonic
crystal cavity:} 
\begin{figure}
\centering
\includegraphics[width=5in]{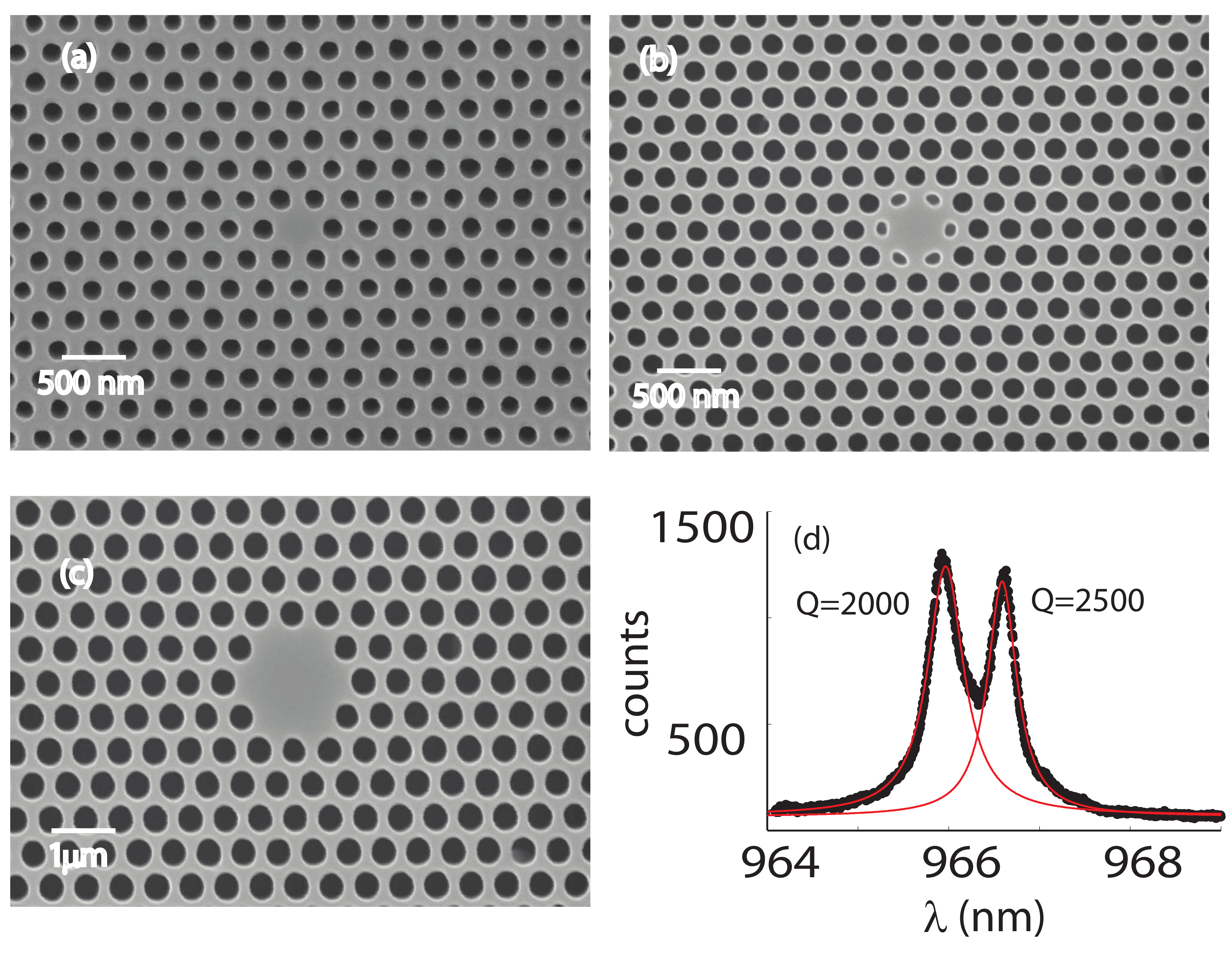}
\caption{Scanning electron micrographs of
fabricated photonic crystal cavities, that can support degenerate
modes: (a) $H1$ cavity, where the central hole is removed; (b)
"calzone" cavity, where the central hole and the half of the next
layer of holes are removed; (c) $H2$ cavity, where the central
hole and the first layer of surrounding holes are removed. (d)
Photoluminescence spectrum of an $H1$ cavity showing moderate
quality factors and moderate degeneracy of the two modes. Back dots correspond to experimental data points; the red lines are Lorentzian fits.} \label{Fig_exp_a}
\end{figure}

In this section, we will describe some initial
experimental results showing signature of QD-cavity coupling in
bimodal cavity, and comment on the feasibility of implementing our
theoretical method in this system. Fig. \ref{Fig_exp_a} a,b,c show
the scanning electron microscope images of three different photonic-crystal cavities
designed to support degenerate cavity modes. Fig
\ref{Fig_exp_a}d shows the photoluminescence (PL) spectrum from an
$H1$ (Fig. \ref{Fig_exp_a} a) cavity. We have so far achieved quality factors of
$2000-2500$ with this cavity design while maintaining
significant overlap between the two modes. For this particular system
the separation between the two peaks is $\sim 0.5$ nm.

\begin{figure}
\centering
\includegraphics[width=5in]{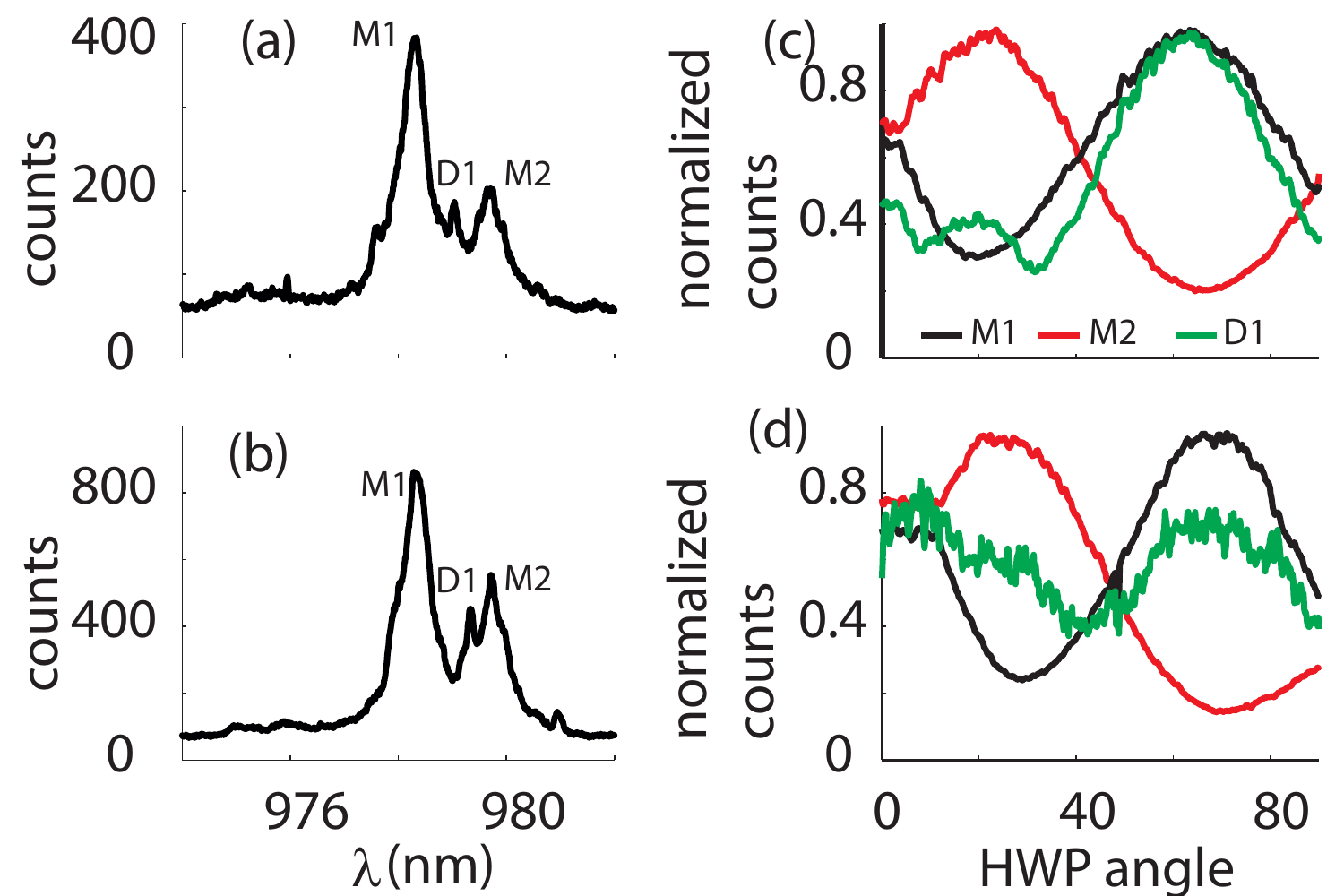}
\caption{(color online) (a), (b) PL spectrum of the bimodal cavity (H2 cavity)
and a coupled QD at two different temperatures $20$K and $30$K. The two modes are denoted as M1 and M2 and the QD is denoted as D1.
(c),(d) Polarization dependence of the cavity modes (red and black plot) and the QD (green plot) at
two different temperatures. } \label{Fig_exp_b}
\end{figure}

Lastly, we study the effects
of the drive laser polarization to verify that  the QD can couple to both cavity modes, and to examine the polarization properties of the two cavity modes.
Fig. \ref {Fig_exp_b}a,b show the PL
spectrum of the bimodal cavity system (mode M1 and M2), with a
coupled QDs (D1) at two different temperature $20$ and $30$K. Fig.
\ref {Fig_exp_b}c,d show the polarization dependence of the two
modes and the QD. At both temperatures, the cavities are of
orthogonal polarizations. However, in Fig. \ref {Fig_exp_b}c the
QD D1, mostly follows the cavity mode M1. With increasing
temperature, the QD moves closer to mode M2, and the polarization
dependence of D1, follows both mode M1 and M2 (Fig. \ref
{Fig_exp_b}d).

Based on these preliminary results, we expect that modest fabrication improvements increasing the frequency overlap of the two modes or or implementation of in-situ frequency tuning of the two modes will allow us to implement the proposed scheme.

\bibliography{Blockade_degen}
\end{document}